\documentclass[prb,amssymb,showkeys]{revtex4}

\newcommand{\be}{\begin{equation}}
\newcommand{\ee}{\end{equation}}
\newcommand{\ba}{\begin{eqnarray}}
\newcommand{\ea}{\end{eqnarray}}

\begin{document}
\title{Kinematics of a Spacetime with an Infinite Cosmological Constant}

\author{R. Aldrovandi}

\author{A. L. Barbosa}

\author{M. Cal\c cada}

\author{J. G. Pereira}
\affiliation{Instituto de F\'{\i}sica Te\'orica \\
Universidade Estadual Paulista\\ Rua Pamplona 145 \\
01405-900 S\~ao Paulo SP \\ Brazil}

\begin{abstract}
A solution of the sourceless Einstein's equation with an infinite value for the cosmological
constant $\Lambda$ is discussed by using In\"on\"u--Wigner contractions of the de Sitter groups
and spaces. When $\Lambda \rightarrow \infty$, spacetime becomes a four--dimensional cone, dual
to Minkowski space by a spacetime inversion. This inversion relates the four--cone vertex to
the infinity of Minkowski space, and the four--cone infinity to the Minkowski light--cone. The
non-relativistic limit $c \rightarrow \infty$ is further considered, the kinematical group in
this case being a modified Galilei group in which the space and time translations are replaced
by the non-relativistic limits of the corresponding proper conformal transformations. This
group presents the same abstract Lie algebra as the Galilei group and can be named the {\it
conformal Galilei} group. The results may be of interest to the early Universe Cosmology.
\end{abstract}


\maketitle

\renewcommand{\thesection}{\arabic{section}}
\section{Introduction}

A kinematical group, whichever it may be, will always have a subgroup accounting for both the
isotropy of space (rotation group) and the equivalence of inertial frames (boosts). The
remaining transformations, generically called {\it translations}, can be either commutative or
not, and are responsible for the homogeneity of space and time. This holds of course for usual
special--relativistic kinematics, but also for Galilean and other conceivable non--relativistic
kinematics,$^($\cite{levy}$^)$ which differ from each other precisely by being grounded on
different kinematical groups.

The best known example is the Poin\-ca\-r\'e group ${\cal P}$, a group naturally associated
to Minkowski spacetime $M$ as its group of motions. It contains, in the form of a semi--direct
product, the Lorentz group ${\cal L} = SO(3,1)$ and the translation group ${\cal T}$. The
latter acts transitively on $M$ and its manifold is just $M$. Indeed, Minkowski  spacetime is a
homogeneous space under ${\cal P}$, actually the quotient $M \equiv {\cal T} = {\cal P}/{\cal
L}$. The invariance of $M$ under the transformations of ${\cal P}$ reflects its uniformity. The
Lorentz subgroup provides an isotropy around a given point of $M$, and translation
invariance enforces this isotropy around any other point.  This is the usual meaning of
``uniformity", in which ${\cal T}$ is responsible for the equivalence of all points of
spacetime.
	
The concept of group contraction, on the other hand, was first
introduced$^($\cite{inonu1,inonu2}$^)$ to formalize and generalize the well known fact that the
Galilei group can be obtained from the  Poin\-ca\-r\'e group in the non-relativistic limit $c
\rightarrow \infty$. The general procedure of group contraction involves always a preliminary
choice of convenient coordinates, in terms of which a certain parameter is made explicit. A new
kinematics can then be obtained by taking this parameter to an appropriate limit. In the
specific case of the contraction of the  Poin\-ca\-r\'e to the Galilei group, the parameter is
the speed of light $c$, and the limit is achieved by taking $c$ to infinity.

Another well known example of group contraction is that by which the  Poin\-ca\-r\'e group is
obtained from any of the two de Sitter groups through a non-cosmological
limit.$^($\cite{gursey}$^)$ In that case, the contraction parameter is the de Sitter
pseudo--radius ${\cal R}$. The contraction is achieved by taking the limit ${\cal R} \rightarrow
\infty$, under which the de Sitter ``translations'' reduce to the Poincar\'e  space and time
translations. The de Sitter spaces are solutions of Einstein's equation for an empty space with
a nonvanishing cosmological constant $\Lambda = R/4$, with $R$ the scalar curvature of the de
Sitter spaces. Since $R \propto {\cal R}^{-2}$ [see Eq.\ (\ref{49}) below], this limit is
equivalent to that in which $\Lambda \rightarrow 0$.

Inflationary cosmological models, on the other hand, suppose a very high value of the
cosmological constant $\Lambda$ at the early stages of the Universe. The relation between its
present time$^($\cite{recent}$^)$ value $\Lambda_0$ and its value $\Lambda_{GUT}$ when the
breakdown of the grand unification symmetry has taken place is$^($\cite{narlikar}$^)$
\be
\Lambda_{GUT} \approx 10^{108} \; \Lambda_0 \; .
\ee
Subsequently, at the breakdown of the electroweak symmetry, there has been another phase
transition with energy--scale of the  order
\be
\Lambda_{EW} \approx 10^{57} \; \Lambda_0 \; .
\ee
Thus, according to this scheme, the cosmological constant was very high at the beginning of
the Universe, changing later to $\Lambda_{GUT}$, then to $\Lambda_{EW}$, and finally to
$\Lambda_0$.  How $\Lambda$ managed to change from a very large initial value to its 
present value is still an open question.

Now, for symmetry reasons, it is appealing to assume an infinite primordial $\Lambda$ at some
initial time, to be followed in the succeeding epochs by a decaying (either through a time
dependence or through phase transitions) but still large cosmological term which would drive
inflation.  All matter (energy) becomes negligible in the presence of an infinite cosmological
constant, so that at that initial stage the Universe spacetime should be a limit solution of
the sourceless Einstein's equation with an infinite $\Lambda$. By using In\"on\"u--Wigner
contractions$^($\cite{inonu1,inonu2}$^)$ of the de Sitter groups and spaces, it has already
been shown that this solution is a four--dimensional cone--spacetime, whose corresponding
kinematical group is the {\it conformal} Poincar\'e group.$^($\cite{ap1}$^)$ On this
cone--spacetime, the metric is singular everywhere  except on a subspace, the
three--dimensional light--cone. Rather surprisingly, this singular character of the metric does
not prevent the existence of well--defined Levi--Civita connection and Riemann curvature
tensor. It makes of the four--dimensional cone, nonetheless, a very strange spacetime. Distance
can be defined only on the light--cone,  where it is zero between any two points. In
particular, no distance is defined on the spacelike sections.  Space sections can, however,
recover a notion of distance in the non--relativistic limit of infinite speed of light $c$. In
that limit, the cone lends its limit metric to the space section. Our aim in this paper will be
to study, by using In\"on\"u--Wigner contractions, the geometrical properties and the
kinematical group of such a speculative spacetime, on which both the cosmological constant and
the speed of light are infinite. We start with a brief review of the de Sitter groups and
spaces.

\section{The de Sitter Groups and Spaces} 
\noindent
The de Sitter spaces are the only possible uniformly curved four--dimensional metric
spacetimes.$^($\cite{weinberg}$^)$ There are two kinds of them,$^($\cite{ellis}$^)$ one with
positive, and one with negative curvature. They can be defined as hypersurfaces in the
pseudo--Euclidean spaces ${\bf E}^{4,1}$ and ${\bf E}^{3,2}$, inclusions whose points in
Cartesian coordinates $(\xi^A) = (\xi^0, \xi^1, \xi^2, \xi^3$, $\xi^{4})$ satisfy,
respectively,
\ba
\eta_{AB} \xi^A \xi^B &\equiv& - (\xi^0)^2 + (\xi^1)^2 + (\xi^2)^2 + (\xi^3)^2 + (\xi^{4})^2 =
{\cal R}^2 \; ; \\  {} \nonumber \\
\eta_{AB} \xi^A \xi^B &\equiv& - (\xi^0)^2 + (\xi^1)^2 + (\xi^2)^2 + (\xi^3)^2 - (\xi^{4})^2 =
- {\cal R}^2 \; .
\ea
We use $\eta_{a b}$ ($a, b = 0,1,2,3$) for the Lorentz metric $\eta = $ diag $(-1$, $1$, $1,
1)$, and the notation $\epsilon = \eta_{44}$ to put both the above conditions together as
\be
\epsilon \; \eta_{a b} \, \xi^{a} \xi^{b} + \left(\xi^4\right)^2 = {\cal R}^2 \; .
\label{dspace2}
\ee
Defining the scaled coordinate $\xi^{\prime 4} = \xi^4/{\cal R}$, one has
\be
\frac{\epsilon}{{\cal R}^2} \; \eta_{a b} \, \xi^{a} \xi^{b} + \left(\xi^{\prime 4}\right)^2 = 1
 \; ,
\label{dspace1}
\ee
where ${\epsilon}/{{\cal R}^2}$ represents the Gaussian curvature.
 
The de Sitter space $dS(4,1)$, whose metric is derived from the pseudo--Euclidean
metric $\eta_{AB}$ = $(-1,+1,+1,+1,+1)$, has the pseudo--orthogonal group $SO(4,1)$ as group of
motions. The other, which comes from $\eta_{AB}$ = $(-1,+1,+1,+1,-1)$, is frequently called
anti--de Sitter space and is denoted $dS(3,2)$ because its group of motions is $SO(3,2)$. The de
Sitter spaces are both homogeneous spaces: 
\begin{equation}
dS(4,1) = SO(4,1)/ SO(3,1) \quad \mbox{and} \quad dS(3,2) = SO(3,2)/ SO(3,1) \;
.\label{eq:quotients}
\end{equation}
The manifold of each de Sitter group is a bundle with the corresponding de Sitter space
as base space and ${\cal L}=SO(3,1)$ as fiber.$^($\cite{kono}$^)$

In the Cartesian coordinates $\xi^A$, the generators of the infinitesimal de Sitter
transformations are given by
\be
J_{A B} = \eta_{AC} \, \xi^C \, \frac{\partial}{\partial \xi^B} -
\eta_{BC} \, \xi^C \, \frac{\partial}{\partial \xi^A} \; ,
\label{dsgene}
\ee
which satisfy the commutation relations
\be
\left[ J_{AB}, J_{CD} \right] = \eta_{BC} J_{AD} + \eta_{AD} J_{BC} - \eta_{BD} J_{AC}
- \eta_{AC} J_{BD} \; .
\label{dsal}
\ee

The four--dimensional stereographic coordinates $x^\mu$ are defined by$^($\cite{gursey}$^)$
\be
\xi^{a} = n(x) \, \delta^{a}{}_{\mu} \, x^\mu \equiv h^{a}{}_{\mu} \, x^\mu \quad
\mbox{and} \quad \xi^4 = - {\cal R} \,  n(x) \left(1 - \epsilon \,
\frac{\sigma^2}{4 {\cal R}^2} \right) \; , 
\label{xix}
\ee
where
\be
n(x) = \frac{1}{1+ \epsilon \sigma^2 / 4 {\cal R}^2} 
\label{n}
\ee
and
\be
\sigma^2 = \eta_{\mu \nu} \, x^\mu x^\nu \; ,
\ee
with $\eta_{\mu \nu} = \delta^{a}{}_{\mu} \, \delta^{b}{}_{\nu} \, \eta_{a b}$.
The $h^{a}{}_{\mu}$ introduced in (\ref{xix}) are the components of a tetrad field,
actually of the 1-form basis members $\omega^{a} = h^{a}{}_{\mu} dx^\mu = n(x) \,
\delta^{a}{}_{\mu} dx^\mu$.

In these coordinates, the line element
\be
ds^2 = \eta_{AB} \, d\xi^A d\xi^B
\ee
is found to be $ds^2 = g_{\mu \nu} \,d x^\mu dx^\nu$, with
\be
g_{\mu \nu} = h^{a}{}_{\mu} \, h^{b}{}_{\nu} \, \eta_{a b} \equiv n^2(x) \, \eta_{\mu \nu}
\label{44}
\ee
the corresponding metric tensor. The de Sitter spaces, therefore, are conformally flat,
with the conformal factor given by $n^2(x)$. We could have written simply $\xi^\mu = n(x) \,
x^\mu$, but we are carefully using the Latin alphabet for the algebra (and flat space)
indices, and the Greek alphabet for the homogeneous space fields and cofields. As usual
with changes from flat tangent--space to spacetime, letters of the two kinds are
interchanged with the help of the tetrad field. This is true for all tensor indices.
Connections, which are vectors only in the last (1-form) index, will gain an extra
``vacuum" term.$^($\cite{livro}$^)$

The Christoffel symbol corresponding to the metric $g_{\mu \nu}$ is
\be
\Gamma^{\lambda}{}_{\mu \nu} = \left[ \delta^{\lambda}{}_{\mu}
\delta^{\sigma}{}_{\nu}  + \delta^{\lambda}{}_{\nu}
\delta^{\sigma}{}_{\mu} - \eta_{\mu \nu} \eta^{\lambda \sigma} \right]
\partial_\sigma [\ln n(x)] \; .
\label{46}
\ee
The corresponding Riemann tensor components, 
\be
R^{\mu}{}_{\nu \rho \sigma} = \partial_\rho
\Gamma^{\mu}{}_{\nu \sigma} - \partial_\sigma \Gamma^{\mu}{}_{\nu \rho} +
\Gamma^{\mu}{}_{\epsilon \rho} \, \Gamma^{\epsilon}{}_{\nu \sigma} -
\Gamma^{\mu}{}_{\epsilon \sigma} \, \Gamma^{\epsilon}{}_{\nu \rho} \; ,
\ee
are found to be
\be
R^{\mu}{}_{\nu \rho \sigma} = \epsilon \, \frac{1}{{\cal R}^2} \,
\left[\delta^{\mu}{}_{\rho} g_{\nu \sigma} - \delta^{\mu}{}_{\sigma} g_{\nu
\rho} \right] \; .
\label{47}
\ee
The Ricci tensor and the scalar curvature are, consequently,
\be
R_{\mu \nu} = \epsilon \, \frac{3}{{\cal R}^2} \, g_{\mu \nu} \quad
{\rm and} \quad
R = \epsilon \, \frac{12}{{\cal R}^2} \;.
\label{49}
\ee
	
In terms of the coordinates $\{x^\mu\}$, the generators (\ref{dsgene}) of the infinitesimal de
Sitter transformations are given by
\ba
J_{a b} &\equiv& \delta_{a}{}^{\mu} \, \delta_{b}{}^{\nu} \,
\left( \eta_{\rho \mu} \, x^\rho \, P_\nu -
\eta_{\rho \nu} \, x^\rho \, P_\mu \right) \label{dslore} \\
{} \nonumber \\
J_{a 4} &\equiv& \epsilon \, \delta_{a}{}^{\mu} \, 
\left({\cal R} \, P_\mu + \frac{\epsilon}{4 {\cal R}} \, K_{\mu} \right) \; ,
\label{dstra}
\ea
where
\be
P_\mu = \frac{\partial}{\partial x^\mu} \quad {\rm and} \quad
K_\mu = \left(2 \eta_{\mu \lambda} x^\lambda x^\rho - \sigma^2 \delta_{\mu}{}^{\rho}
\right) P_\rho
\label{cp2} 
\ee
are respectively the generators of translations and proper conformal transformations.
For $\epsilon = +1$, we get the generators of the de Sitter group $SO(4,1)$. For
$\epsilon = -1$, we get the generators of the de Sitter group $SO(3,2)$. The $J_{a
4}$'s of (\ref{dstra}) behave as translation generators on the corresponding homogeneous
spaces, while the $J_{a b}$'s span the Lorentz subgroup $SO(3,1)$. 

Notice that the above
expressions for the generators involve quantities appearing also in the metric and the
curvature. Geometry and algebra are deeply mixed, as  a result of the quotient character
(\ref{eq:quotients}) of spacetime. De Sitter spacetimes are actually imbedded in the
manifolds of the de Sitter groups. We shall now proceed to deform the algebras and,
consequently, the groups. The imbedded spacetimes will follow these deformations, changing
accordingly.

\section{Infinite Cosmological--Constant Contraction}
 \label{sec:lambdacontraction}
\noindent	
Let us consider now the limit ${\cal R} \rightarrow 0$. First, we rewrite Eqs.\
(\ref{dslore}) and (\ref{dstra}) in the form
\ba
J_{a b} &\equiv& \delta_{a}{}^{\mu} \, \delta_{b}{}^{\nu} L_{\mu \nu} \\
{} \nonumber \\
J_{a 4} &\equiv& {\cal R}^{-1} \, \delta_{a}{}^{\mu} \, \kappa_{\mu}  \; ,
\ea
where
\be
L_{\mu \nu} =
\left( \eta_{\rho \mu} \, x^\rho \, P_\nu -
\eta_{\rho \nu} \, x^\rho \, P_\mu \right)
\label{cp1}
\ee
are the generators of the Lorentz group, and
\be
\kappa_{\mu} = {\textstyle {\frac{1}{4}}} \, K_\mu + \epsilon \, {\cal
R}^2 \, P_{\mu} \; .
\label{pmuz}
\ee
In terms of these generators, the commutation relations (\ref{dsal}) become
\ba
\left[L_{\mu \nu}, L_{\lambda \rho}\right] &=& \eta_{\nu \lambda} \,
L_{\mu \rho} + \eta_{\mu \rho} \, L_{\nu \lambda} - \eta_{\nu \rho} \,
L_{\mu \lambda} - \eta_{\mu \lambda} \, L_{\nu \rho} \; , \label{llz} \\
{} \nonumber \\
\left[\kappa_{\mu},L_{\lambda \rho}\right] &=& \eta_{\mu \lambda} \,
\kappa_{\rho} - \eta_{\mu \rho} \, \kappa_{\lambda} \; , \label{plz} \\
{} \nonumber \\
\left[\kappa_{\mu}, \kappa_{\lambda}\right] &=& - {\cal R}^{2}
\, L_{\mu \lambda} \; .
\label{ppz}
\ea

Now, in the contraction limit ${\cal R} \rightarrow 0$, one can see that
\be
\lim_{{\cal R}\to 0} \,L_{\mu \nu} = L_{\mu \nu} \quad ; \quad
\lim_{{\cal R}\to 0} \, \kappa_\mu = {\textstyle {\frac{1}{4}}} \, K_\mu \;
,
\ee
and consequently the de Sitter algebra contracts to 
\ba
\left[L_{\mu \nu}, L_{\lambda \rho}\right] &=& \eta_{\nu \lambda} \,
L_{\mu \rho} + \eta_{\mu \rho} \, L_{\nu \lambda} - \eta_{\nu \rho} \,
L_{\mu \lambda} - \eta_{\mu \lambda} \, L_{\nu \rho} \; , \label{llx} \\
{} \nonumber \\
\left[ K_{\mu}, L_{\lambda \rho}\right]&=&\eta_{\mu \lambda}  K_{\rho} -
\eta_{\mu \rho}  K_{\lambda} \; , \label{klx} \\
{} \nonumber \\
\left[ K_{\mu},  K_{\lambda}\right]&=&0 \; .
\label{kk}
\ea
These commutation relations coincide with those of the Poincar\'e group Lie
algebra. However, the Lie group corresponding to this algebra, denoted by ${\cal Q}$ and formed
by a semi--direct product of Lorentz and proper conformal transformations, is completely
different from the usual Poincar\'e group.  It is the group ruling the
local kinematics of high--$\Lambda$ spaces and has been called the {\it
second} or {\it conformal} Poincar\'e group$^($\cite{ap1}$^)$. Though the first terminology may
have more appeal to a physicist,  the second is more precise$^($\cite{NORM97}$^)$. 

From Eq.\ (\ref{dspace1}) one sees that in the limit ${\cal R} \to \infty$, $(\xi'^4)^2 =
1$, the curvature vanishes, and one obtains the flat Minkowski space, a
$4$-dimensional hyperplane in the  $5$-dimensional linear ambient space. In terms of the
variable $\xi^4$, Eq.\ (\ref{dspace2}) shows that the contraction limit ${\cal R} \rightarrow
0$ leads both de Sitter spaces to a four--dimensional cone--space, denoted by $N$, in which $ds
= 0$. It presents a geometry gravitationally related to an infinite cosmological constant, and
its kinematical group of motions is ${\cal Q}$. Analogously to the Minkowski case, $N$ is also
a homogeneous space, but under the kinematical group ${\cal Q}$, that is, $N = {\cal Q}/{\cal
L}$. The point--set of $N$ is the point--set of the proper conformal transformations. The
kinematical group ${\cal Q}$, like the Poincar\'e group, has the Lorentz group ${\cal L}$ as
the subgroup accounting for the isotropy of the cone--space $N$. However, the proper conformal
transformations introduce a new kind of homogeneity. In fact, instead of ordinary translations,
all points of $N$ are equivalent through proper conformal transformations.

An important property of the cone--space $N$ is that its metric tensor is singular everywhere,
\be
\lim_{{\cal R}\to 0} \, g_{\mu \nu} = 0 \; ; \quad
\lim_{{\cal R}\to 0} \, g^{\mu \nu} \to \infty \; , 
\ee
except on the points defined by $\sigma^2 = 0$, where
\be
g_{\mu \nu} = \eta_{\mu \nu} \; .
\ee
In other words, the metric turns out to be defined only on the three--dimensional light--cone
subspace of the cone--spacetime $N$. It is singular at every other point. Nevertheless, the
Levi--Civita connection (\ref{46}) is well defined everywhere, and consequently the Riemann
curvature tensor is also well defined. An explicit computation shows that, whereas both the
Riemann and the Ricci curvature tensors vanish for ${\cal R} \to 0$, the scalar
curvature goes to infinity:
\be
\lim_{{\cal R} \to 0} \, R \to \infty \; .
\ee
This is a characteristic property of a spacetime with an infinite cosmological constant.
Finally, it is important to mention that the {\it conformal} Poincar\'e group ${\cal Q}$,
which is the group of motion of the cone--space $N$, preserves the light--cone structure.

\section{Infinite Speed of Light Contraction}
\noindent
As already mentioned, no distance can be defined on the space--like sections of the cone--space
$N$. However, a notion of distance can be recovered in the non--relativistic limit. 
Let us then proceed  to examine the limit in which the speed of light $c$ goes to infinity. It
has been  emphasized  by Bacry \& L\'evy-Leblond$^($\cite{levy}$^)$ that this limit corresponds
to a situation in which not only velocities are small, but also that spacelike intervals are
small as compared to timelike intervals. In order to perform this contraction, we need to
introduce new coordinates so as to exhibit $c$ explicitly.$^($\cite{inonu1}$^)$ Denoting the
old coordinates with a bar, we define new coordinates
$x^\mu$ according to
\be
\bar{x}^\mu = \frac{1}{c} \, x^\mu \; .
\ee
In terms of the new coordinates, the generators $L_{\mu \nu}$ and $K_\mu$ of the conformal
Poinca\-r\'e group, given respectively by Eqs.\ (\ref{cp1}) and (\ref{cp2}), become $(i=1,2,3)$
\ba
L_{ij} &=& \eta_{ik} x^k P_j - \eta_{jk} x^k P_i \\
{} \nonumber \\
L_{i4} &=& - c B_i \\
{} \nonumber \\
K_i &=& c T_i \\
{} \nonumber \\
K_4 &=& T_{t} \; ,
\ea
where we have used the notation
\ba
B_i &=& t P_i - \eta_{ik} \, \frac{x^k}{c^2} \, P_{t} \\
{} \nonumber \\
T_i &=& 2 \eta_{ik} \, \frac{x^k x^j}{c^2} \, P_j +
2 \eta_{ik} \, \frac{x^k t}{c^2} \, P_{t} - t^2 P_i + \frac{r^2}{c^2} \, P_i \\
{} \nonumber \\
T_{t} &=& 2 t x^i P_i + t^2 P_{t} + \frac{r^2}{c^2} \, P_{t}
\ea
with
$P_{t} = {\partial}/{\partial t}$ and $P_i = {\partial}/{\partial x^i}$.
Notice furthermore that
\be
\sigma^2 = - t^2 + (r^2/c^2) \; , 
\ee
with $r^2=(x^1)^2 + (x^2)^2 + (x^3)^2$.
In terms of the generators $L_{ij}$, $B_i$, $T_i$ and $T_{t}$, the commutation relations
(\ref{llx}), (\ref{klx}) and (\ref{kk}) of the conformal Poincar\'e group can be rewritten in
the form:
\ba
\left[L_{i j}, L_{k l}\right] &=& \eta_{j k} \,
L_{i l} + \eta_{i l} \, L_{j k} - \eta_{j l} \,
L_{i k} - \eta_{i k} \, L_{j l} \\
{} \nonumber \\
\left[L_{i j}, B_{k}\right] &=& \eta_{j k} \, B_i - \eta_{i k} \, B_j \\
{} \nonumber \\
\left[B_{i}, B_{k}\right] &=& - \frac{1}{c^2} \, L_{i k} \\
{} \nonumber \\
\left[T_{i}, B_{k}\right] &=& - \frac{1}{c^2} \, \eta_{i k} T_{t} \\
{} \nonumber \\
\left[T_{i}, L_{k l}\right] &=& \eta_{i k} \,
T_{l} - \eta_{i l} \, T_{k} \\
{} \nonumber \\
\left[T_{t}, B_{k}\right] &=& T_k \\
{} \nonumber \\ 
\left[T_{t}, L_{k l}\right] &=& \left[T_i, T_{k}\right] =
\left[T_{t}, T_{t} \right] = 0 \; .
\ea

Let us now consider the limit $c \rightarrow \infty$. It is easy to see that, in this
limit, the generators assume the form
\ba
L_{ij} &=& \eta_{ik} x^k P_j - \eta_{jk} x^k P_i \label{g1} \\
{} \nonumber \\
B_i &=& t P_i \label{g2} \\
{} \nonumber \\
T_i &=& - t^2 P_i \label{g3} \\
{} \nonumber \\
T_{t} &=& 2 t x^i P_i + t^2 P_{t} \label{g4} \; .
\ea
The corresponding commutation relations become
\ba
\left[L_{i j}, L_{k l}\right] &=& \eta_{j k} \,
L_{i l} + \eta_{i l} \, L_{j k} - \eta_{j l} \,
L_{i k} - \eta_{i k} \, L_{j l} \\
{} \nonumber \\
\left[L_{i j}, B_{k}\right] &=& \eta_{j k} \, B_i - \eta_{i k} \, B_j \\
{} \nonumber \\
\left[B_{i}, B_{k}\right] &=&
\left[T_{i}, B_{k}\right] = 0 \\
{} \nonumber \\
\left[T_{i}, L_{k l}\right] &=& \eta_{i k} \,
T_{l} - \eta_{i l} \, T_{k} \\
{} \nonumber \\
\left[T_{t}, B_{k}\right] &=& T_k \\
{} \nonumber \\ 
\left[T_{t}, L_{k l}\right] &=& \left[T_i, T_{k}\right] =
\left[T_{t}, T_{t} \right] = 0 \; .
\ea
This commutation table coincides with the Lie algebra of the Galilei group.
The group, however, is quite distinct from Galilei. The rotation and boost generators, given
respectively by $L_{i j}$ and $B_{k}$, are the same as those of the Galilei group. This means
that the concept of isotropy of space and the equivalence of inertial frames coincide with
that of the Galilei group. Nevertheless, the concepts of time and space homogeneity are
completely different. Instead of ordinary time and space translations, homogeneity in
space and time are defined respectively by the generators $T_i$ and $T_{t}$,  given by
the non-relativistic limit of the proper conformal generators. 
 This new group does deserve, for this reason, the name {\it conformal Galilei} group.

\section{Final Remarks}
\noindent
As is well known, by the process of In\"on\"u--Wigner group contraction with
${\cal R} \rightarrow \infty$, both de Sitter groups are reduced to the Poincar\'e group
${\cal P}$, and both de Sitter spacetimes are reduced to the Minkowski space $M$. On the
other hand, in a similar fashion but taking the limit ${\cal R} \rightarrow 0$, which
corresponds to ${\Lambda} \rightarrow \infty$, both de Sitter groups are contracted to
the {\it conformal} Poincar\'e group ${\cal Q}$, formed by the semi--direct product of
Lorentz and proper conformal transformations, and both de Sitter spaces are reduced to the
cone--space $N$, a spacetime characterized by presenting vanishing Riemann and Ricci
curvature tensors, but an infinite scalar curvature.

Minkowski space and the cone--spacetime can be considered as {\it dual} to each other in the
sense that their geometries are determined, respectively, by a vanishing and an infinite
cosmological constant. The same can be said of their kinematical group of motions: ${\cal P}$
is associate to a vanishing cosmological constant, and ${\cal Q}$ to an infinite cosmological
constant. The {\em duality} transformation connecting these two geometries is the spacetime
inversion
\be
x^\mu \rightarrow - \frac{x^\mu}{\sigma^2} \; .
\ee
Under such a transformation, the Poincar\'e group ${\cal P}$ is transformed into the
{\it conformal} Poincar\'e group ${\cal Q}$, and the Minkowski space $M$ becomes the
four--dimensional cone--space $N$. The points at infinity of $M$ are concentrated in the
vertex of the cone--space $N$, and those on the light--cone of $M$ becomes the infinity of $N$.

Now, as we have seen, the metric of $N$ is singular everywhere except on the
three--dimensional light--cone, where it coincides with the Minkowski metric $\eta_{\mu \nu}$.
Although conceivable  from the mathematical point of view, a Universe described by such a
spacetime would be quite peculiar. No distance could be defined, except on the light--cone
where it would be always vanishing. However, in the non--relativistic limit $c \rightarrow
\infty$, the notion of distance is recovered. On the other hand, as is widely known, under this
limit, the Poincar\'e group is contracted to the Galilei group, and the Minkowski spacetime is
divided into two disconnected pieces, a three--dimensional Euclidean space and time, which
becomes a parameter. Under the same limit, the {\it conformal} Poincar\'e group ${\cal Q}$ is
contracted to a group that includes the same three--dimensional rotation and boost
transformations of the Galilei group, but time and space translations are replaced by the
corresponding non-relativistic limit of the proper conformal transformation. Interesting enough,
this group presents the same abstract Lie algebra as the Galilei group, and for this reason it
has been named the {\it conformal} Galilei group. Analogously to what occurs to the Minkowski
space, for $c \rightarrow \infty$ the cone--space $N$ is divided into two disconnected parts, a
three--dimensional Euclidean space and time, which also in this limit becomes a parameter.
Furthermore, as the three--dimensional Euclidean space comes from the light--cone, where the
metric is well defined, the metric of the resulting Euclidean space will be well defined also.
The group of motions of this space is the conformal Galilei group, whose generators of
infinitesimal transformations are those given by Eqs.~(\ref{g1})-(\ref{g4}).

It is important to mention finally that the order of the contractions
($\Lambda \rightarrow \infty$ then $c \rightarrow \infty$, or $c \rightarrow \infty$ then
$\Lambda \rightarrow \infty$) is not important for the results obtained. The intermediary
results, however, would change. In fact, the non-relativistic limit ($c \rightarrow \infty$)
of the de Sitter groups and spacetimes leads respectively to the Newton--Hooke group and
spacetime.$^($\cite{abcp}$^)$ A further infinite cosmological contraction ($\Lambda \rightarrow
\infty$) would then lead to the conformal Galilei group. 

\begin{acknowledgments}
The authors would like to thank FAPESP-Brazil and CNPq-Brazil for financial support. They 
 thank also   two unknown referees whose suggestions helped to improve the text.
\end{acknowledgments}

\end{document}